\documentclass[twocolumn,showpacs,floatfix,prl]{revtex4}

\usepackage[dvips]{epsfig}

\usepackage{float}

\begin{document}

\title{Quantum Spin Hall Effect in Graphene}

\author{C.L. Kane and E.J. Mele}
\affiliation{Dept. of Physics and Astronomy, University of Pennsylvania,
Philadelphia, PA 19104}

\begin{abstract}
We study the effects of spin orbit interactions on the low energy
electronic structure of a single plane of graphene. We find that in an
experimentally accessible low temperature regime the symmetry allowed spin
orbit potential converts graphene from an ideal two dimensional
semimetallic state to a quantum spin Hall insulator. This novel  electronic state of
matter is gapped in the bulk and supports the  transport of spin
and charge in gapless
edge states that propagate at the sample boundaries.
The edge states are non chiral, but they are insensitive to disorder
because their directionality is correlated with spin.
The spin and charge
conductances in these edge states are calculated and the effects of
temperature, chemical potential, Rashba coupling, disorder and symmetry breaking
fields are discussed.
\end{abstract}

\maketitle

The generation of  spin currents solid state systems has been a
focus of intense recent interest.  It has been argued that in
doped semiconductors the spin orbit (SO) interaction leads to a
spin-Hall effect\cite{murakami1,sinova}, in which a spin current
flows perpendicular to an applied electric field.
The spin Hall effect has been observed in
GaAs\cite{kato,otherexpt}.
Murakami et
al. \cite{murakami2} have identified a  class of cubic materials
that are insulators, but nonetheless exhibit a finite spin
Hall conductivity.   Such a ``spin Hall insulator"
would be of intrinsic interest, since it would allow for spin
currents to be generated without dissipation.

In this paper we show that at sufficiently low energy a single
plane of graphene exhibits a quantum spin-Hall (QSH) effect with an
energy gap that is generated by the SO interaction. Our
motivation is twofold. First, Novoselov et al.\cite{novoselov}
have recently reported progress in the preparation of single
layer graphene films.  These films exhibit the expected ambipolar
behavior when gated and have mobilities up to $10^4 {\rm
cm}^2/{\rm Vs}$. Thus, the detailed experimental study of
graphene now appears feasible.  We believe the QSH effect
in graphene is observable below a low but experimentally
accessible temperature. Secondly, we will show the QSH
effect in graphene is different from the spin hall effects
studied for three dimensional cubic systems in Ref.
\onlinecite{murakami2} because it leads to a phase which is
{\it topologically distinct} from a band insulator. The QSH effect in graphene
resembles the charge quantum Hall effect, and we
will show that spin and charge currents can be transported in
gapless edge states.  As a model system, graphene thus identifies
a new class of spin Hall insulator. It may provide a starting
point for the search for other spin-Hall insulators in two
dimensional or in layered materials with stronger SO
interaction.

SO effects in graphite have been known for over 40 years
\cite{dresselhaus}, and play a role in the formation of minority
hole pockets in the graphite Fermi surface\cite{review}. However,
these effects have largely been ignored because they are
predicted to be quite small and they are overwhelmed by the
larger effect of coupling between the graphene
planes.  Unlike graphite which has a finite Fermi surface,
however, graphene is in a critical electronic state which can be
strongly affected by small perturbations at low energy.

Graphene consists of a honeycomb lattice
of carbon atoms with two sublattices.  The states near the Fermi
energy are $\pi$ orbitals residing near the $K$
and $K'$ points at opposite corners of the hexagonal Brillouin
zone.  An effective mass model can be developed \cite{ddv} by
writing the low energy electronic wavefunctions as
\begin{equation}
\Psi({\bf r}) =  [(u_{AK}, u_{BK}),(u_{AK'},u_{BK'})]\cdot \psi({\bf r})
\end{equation}
where $u_{(A,B)(K,K')}({\bf r})$ describe basis states at
momentum $k=K$, $K'$ centered on atoms of the $A$, $B$
sublattice. $\psi({\bf r})$ is a four component slowly varying
envelope function.  The effective mass Hamiltonian then takes the
form,
\begin{equation}
{\cal H}_0 = -i\hbar v_F \psi^\dagger(\sigma_x\tau_z\partial_x +
\sigma_y\partial_y)\psi.
\end{equation}
Here $\vec\sigma$ and $\vec\tau$ are Pauli matrices with
$\sigma_z = \pm 1$ describing states on the $A(B)$ sublattice and
$\tau_z = \pm 1$ describing states at the $K(K')$ points.  This
Hamiltonian describes gapless states with $E({\bf q}) = \pm v_F
|{\bf q}|$. Without spin, the degeneracy at ${\bf q}=0$ is
protected by symmetry. The only possible terms that could be
added to open a gap are proportional to $\sigma_z$ or
$\sigma_z\tau_z$.  The $\sigma_z$ term, which corresponds to a
staggered sublattice potential is odd under parity (which
interchanges the A and B sublattices).  The $\sigma_z\tau_z$ term
is even under parity, but odd under time reversal (which
interchanges $K$ and $K'$).

The SO interaction allows for a new term, which will be the focus
of this paper:
\begin{equation}
{\cal H}_{SO} = \Delta_{so} \psi^\dagger \sigma_z\tau_z s_z \psi.
\end{equation}
Here $s_z$ is a Pauli matrix representing the electron's spin.
This term respects all of the symmetries of graphene, and will be
present.  Below we will explicitly construct this term
from the microscopic SO interaction and estimate its magnitude.
If the mirror symmetry about the plane is
preserved then this is the only allowed spin dependent term at ${\bf
q}=0$.  If the mirror symmetry is
broken (either by a perpendicular electric field or by interaction with
a substrate) then a Rashba term\cite{rashba}
 of the form $({\bf s}\times {\bf p})\cdot\hat z$
is allowed,
\begin{equation}
{\cal H}_R = \lambda_R \psi^\dagger (\sigma_x\tau_z s_y - \sigma_y
s_x)\psi.
\end{equation}
For $\lambda_R=0$, $\Delta_{so}$ leads to an energy gap $2\Delta_{so}$ with $E({\bf
q}) = \pm\sqrt{(\hbar v_F q)^2 + \Delta_{so}^2}$.  For
$0<\lambda_R<\Delta_{so}$ the energy gap $2(\Delta_{so}-\lambda_R)$
remains finite.  For $\lambda_R>\Delta_{so}$ the gap closes, and the
electronic structure is that of a zero gap semiconductor with
quadradically dispersing bands.  In the following we will assume that
$\lambda_R<\Delta_{so}$ and analyze the properties of the resulting
gapped phase.  This assumption is justified by numerical estimates
given at the end of the paper.

The  gap generated by $\sigma_z\tau_z s_z$ is different from the gap that
would be generated by the staggered sublattice potentials,
$\sigma_z$ or $\sigma_z s_z$.  The ground states in the presence
of the latter terms are adiabatically connected to simple
insulating phases at strong coupling where the two sublattices
are decoupled. In contrast, the gap parameter  $\sigma_z \tau_z
s_z$ produces gaps with {\it opposite signs} at the $K$ and $K'$
points. This has no simple strong coupling limit. To connect
smoothly between the states generated by $\sigma_z$ and
$\sigma_z\tau_z s_z$ one must pass through a critical point where
the gap vanishes,
separating ground states with distinct topological orders.

The interaction (3) is related to a model introduced by Haldane\cite{haldane1}
 as a realization
of the parity anomaly in (2+1) dimensional relativistic field
theory. Taken separately, the Hamiltonians for the $s_z=\pm 1$
spins violate time reversal symmetry and are equivalent to
Haldane's model for spinless electrons, which could be realized
by introducing a periodic magnetic field with no net flux.  As
Haldane showed, this gives rise to a $\sigma_z\tau_z$ gap, which
has opposite signs at the $K$ and $K'$ points.  At temperatures
well below the energy gap this leads to a quantized Hall
conductance $\sigma_{xy} = \pm e^2/h$.  This Hall
conductance computed by the Kubo formula can be interpreted as the
topological Chern number
induced by the Berry's curvature in momentum
space\cite{thouless,haldane2}. Since the signs of the gaps in (3)
are opposite for opposite spins, an electric field will induce
opposite currents for the opposite spins, leading to a spin
current ${\bf J}_s = (\hbar/2e)({\bf J}_\uparrow-{\bf
J}_\downarrow)$ characterized by a quantized spin Hall
conductivity
\begin{equation}
\sigma^s_{xy}={e\over{2\pi}}.
\end{equation}
Since spin currents do not couple to experimental probes it is difficult
to directly measure (5).  Moreover, the conservation of $s_z$ will be violated
by the Rashba term (4) as well as terms which couple the $\pi$ and
$\sigma$ orbitals.  Nonetheless, Murakami et al. \cite{murakami3} have
defined a conserved spin $s_{z(c)}$, allowing $\sigma_{xy}^s$ to be computed
via the Kubo formula.  We find that $\sigma_{xy}^s$ computed in this way
is not quantized when $\lambda_R\ne 0$, though the correction
to (5) is small due to carbon's weak SO interaction.

In the quantum Hall effect the bulk topological order requires the
presence of gapless edge states.  We now show that gapless edge
states are also present in graphene.
We will begin by establishing the edge states for $\lambda_R=0$.  We
will then argue that the gapless edge states persist even when $\lambda_R \ne
0$, and that they are robust against weak electron electron interactions and
disorder.  Thus, in spite of the violation of (5) the gapless edge states
characterize a state which is distinct from an ordinary insulator.
This QSH state is different from the insulators discussed in
Ref. \onlinecite{murakami2}, which do not have edge states.  It is also distinct from
the spin Hall effect in doped GaAs, which does not have an energy gap.

For $\lambda_R=0$, the Hamiltonian (2,3) conserves
$s_z$, and the gapless edge states follow from
Laughlin's argument\cite{laughlin}.
Consider a large cylinder
(larger than $\hbar v_F/\Delta_{so}$) and adiabatically insert a quantum
$\phi = h/e$ of magnetic flux quantum down the cylinder (slower than $\Delta_{so}/\hbar$).
The resulting azimuthal Faraday
electric field induces a spin current such that spin $\hbar$ is
transported from one end of the cylinder to the other.  Since an
adiabatic change in the magnetic field cannot excite a particle
across the energy gap $\Delta_{so}$ it follows that there must be gapless
states at each end to accommodate the extra spin.

An explicit description of the edge states requires a model that
gives the energy bands throughout the entire Brillouin zone.
Following Haldane\cite{haldane1}, we introduce a second neighbor tight
binding model,
\begin{equation}
{\cal H} =  \sum_{\langle ij\rangle \alpha}  t c_{i\alpha}^\dagger c_{j\alpha}
+ \sum_{\langle\langle ij \rangle\rangle \alpha\beta} i t_2
\nu_{ij}s^z_{\alpha\beta}  c_{i\alpha}^\dagger
c_{j\beta}.
\end{equation}
The first term is the usual nearest neighbor hopping term.  The
second term connects second neighbors with a spin dependent
amplitude.  $\nu_{ij}=-\nu_{ji} = \pm 1$, depending on the
orientation of the two nearest neighbor bonds ${\bf d}_1$ and ${\bf d}_2$
the electron traverses in going from site $j$ to $i$.  $\nu_{ij}=\pm 1$
if the electron makes a left (right) turn to get to the second bond.
The spin dependent term can be written in a coordinate
independent representation as
 $i( {\bf d}_1\times {\bf d}_2) \cdot{\bf s}$.
At low energy (6)
reduces to (2,3) with $\Delta_{so} = 3\sqrt{3} t_2$.

The edge states can be seen by solving
(7) in a strip geometry.  Fig. 1 shows the
one dimensional energy bands for a strip where the edges are
along the zig-zag direction in the graphene plane.  The bulk
bandgaps at the one dimensional projections of the $K$ and $K'$
points are clearly seen. In addition two
bands traverse the gap, connecting the $K$ and $K'$ points.
These bands are localized at the edges of the strip, and each
band has degenerate copies for each edge.  The edge states are
not chiral since each edge has states which propagate in both
directions.  However, as illustrated in
Fig. 2 the edge states are ``spin filtered" in the sense that
electrons with opposite spin propagate in {\it opposite} directions.
Similar edge states occur for
armchair edges, though in that case the 1D projections of $K$ and
$K'$ are both at $k=0$.
It is interesting to note that for zig-zag edges the edge states persist
for $\Delta_{so}\rightarrow 0$, where they become perfectly
flat\cite{zigzagedge}.  This leads to an enhanced density of states at the Fermi
energy associated with zig-zag edges.  This has
been recently seen in scanning tunneling spectroscopy of graphite
surfaces\cite{stsexpt}.

\begin{figure}
 \centerline{ \epsfig{figure=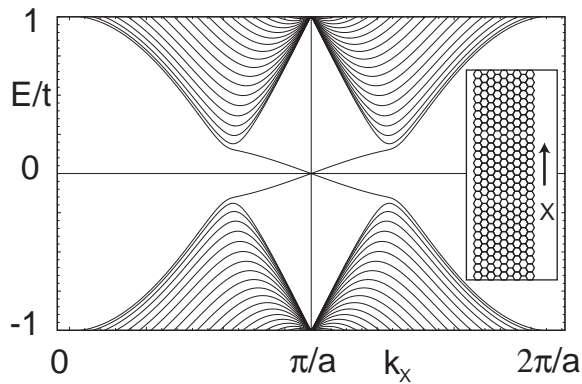,width=3in} }
 \caption{(a) One dimensional energy bands for a strip of graphene (shown in inset)
 modeled by (7) with $t_2/t=.03$.  The bands crossing the gap are spin filtered
  edge states.}
 \end{figure}

We have also considered a nearest neighbor Rashba term, of the form
$i\hat z\cdot({\bf s}_{\alpha\beta}\times{\bf d})c_{i\alpha}^\dagger
c_{j\beta}$.  This violates the conservation of $s_z$, so that the Laughlin
argument no longer applies.  Nonetheless, we find that the gapless
edge states remain, provided $\lambda_R<\Delta_{so}$, so that the bulk
bandgap remains intact.
The crossing of the edge states at the Brillouin zone boundary $k_x = \pi/a$ in
Fig. 1 (or at $k=0$ for the armchair edge) is protected by time
reversal symmetry.  The two states at $k_x=\pi/a$ form a Kramers
doublet whose degeneracy cannot be lifted by any time reversal
symmetric perturbation.  Moreover, the degenerate
states at $k_x = \pi/a \pm q$ are a Kramers doublet.  This means that
elastic backscattering from a random potential is forbidden.
More generally, scattering from a region of disorder can be characterized by a $2\times
2$ unitary S-matrix which relates the incoming and outgoing states:
$\Phi_{\rm out} = S \Phi_{\rm in}$, where $\Phi$ is a two component
spinor consisting of the left and right moving edge states
$\phi_{L\uparrow}$, $\phi_{R\downarrow}$.  Under time reversal
$\Phi_{\rm in,out} \rightarrow s_y \Phi^*_{\rm out,in}$.
Time reversal symmetry therefore imposes the constraint $S = s_y S^T s_y$, which
 rules out any off diagonal elements.

Electron interactions can lead to backscattering.
For instance, the term
$u \psi_{L\uparrow}^\dagger\partial_x \psi_{L\uparrow}^\dagger
\psi_{R\downarrow}\partial_x\psi_{R\downarrow}$,
does not violate time reversal, and will
be present in an interacting Hamiltonian.  For weak
interactions this term is {\it irrelevant} under the renormalization
group, since its scaling
dimension is $\Delta = 4$.  It thus will not lead to an
energy gap or to localization.  Nonetheless, it allows
inelastic backscattering.  To leading order in $u$ it gives a
finite {\it conductivity} of the edge states, which diverges
at low temperature as $u^{-2}T^{3-2\Delta}$\cite{giamarchi}.  Since
elastic backscattering is prevented by time reversal there
are no relevant backscattering processes for weak interactions.
This stability against interactions and disorder
distinguishes the spin filtered edge states from ordinary
one dimensional wires, which are localized by weak disorder.

A parallel magnetic field $H_\parallel$ breaks time reversal and leads to
an avoided crossing of the edge states.  $H_\parallel$ also reduces
the symmetry, allowing
terms in the Hamiltonian which provide a continuously gapped path
connecting the states generated by $\sigma_z\tau_z s_z$ and $\sigma_z$.
Thus in addition to gapping the edge states
$H_\parallel$ eliminates the topological distinction between the QSH phase
and a simple insulator.

\begin{figure}
 \centerline{ \epsfig{figure=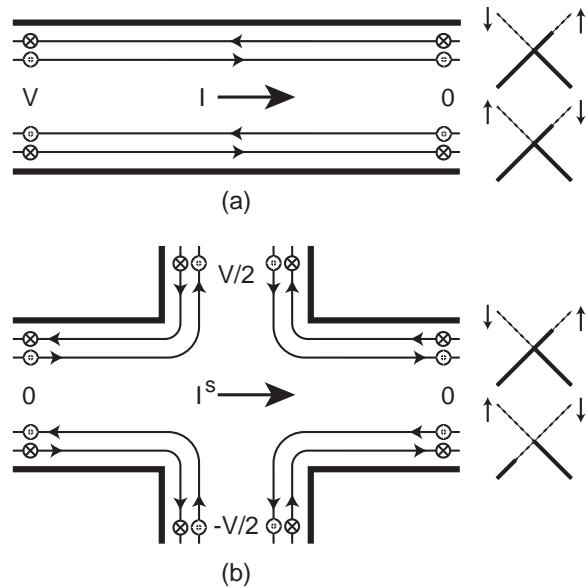,width=3in} }
 \caption{Schematic diagrams showing (a) two terminal and (b) four terminal measurement
 geometries.  In (a) a charge current $I=(2e^2/h)V$ flows into the right lead.
 In (b) a spin current $I^s=(e/4\pi)V$ flows into the right lead.  The diagrams to
 the right indicate the population of the edge states.}
 \end{figure}

The spin filtered edge states have important consequences for
both the transport of charge and spin. In the limit of low temperature
we may ignore the inelastic
backscattering processes, and describe the ballistic transport in the
edge states within a Landauer-B\"uttiker\cite{buettiker} framework.
For a two terminal geometry (Fig. 2a), we predict a ballistic two terminal charge
conductance $G = 2 e^2/h$.  For the spin filtered edge states
the edge current density is related to the spin density, since both
depend on $n_{R\uparrow}-n_{L\downarrow}$.
Thus the charge current is accompanied by spin accumulation at the edges.
The interplay between charge and spin can be probed in a
multiterminal device.  Define the multiterminal
spin conductance by $I^s_i = \sum_j
G^s_{ij}V_j$.  Time reversal symmetry requires
$G^s_{ji} = - G^s_{ij}$, and from Fig. 2b it is clear that
$G^s_{ij} = \pm e/4\pi$ for adjacent contacts $i$ and $j$.
In the four terminal geometry of Fig. 2b a spin  current $I^s = eV/4\pi$
flows into the right contact.  This geometry can also be used to {\it
measure} a spin current.  A spin current incident from the left (injected, for
instance with a ferromagnetic contact) will
be split, with the up (down) spins transported to the top (bottom
contacts), generating a measurable spin-Hall voltage.

The magnitude of $\Delta_{so}$ may be estimated
by treating the microsopic SO interaction
\begin{equation}
V_{SO}={\hbar\over{4 m^2 c^2}}{\bf s} \cdot({\bf \nabla}V \times {\bf p})
\end{equation}
 in first order degenerate
perturbation theory.  We thus evaluate the expectation value of
(8) in the basis of states given in (1) treating  $\psi({\bf r})$ as a constant.
A full evaluation depends on the detailed form of the Bloch functions.
However a simple estimate can be made in the ``first
star" approximation: $u_{(K,K'),(A,B)}({\bf
r}) = \sum_p \exp [ i {\bf K}_p\cdot({\bf r} - {\bf d})]
/\sqrt{3}$. Here ${\bf K}_p$ are the crystal momenta at the three
corners of the Brillouin zone equivalent to $K$ or $K'$, and
${\bf d}$ is the a
basis vector from a hexagon center to an A or B sublattice site.
We find that the matrix elements have precisely the structure (3), and
using the Coulomb interaction $V(r) = e^2/r$ we estimate $2\Delta_{so} =
4\pi^2 e^2 \hbar^2 /(3 m^2 c^2 a^3) \sim 2.4^\circ K$. This is a
crude estimate, but it is comparable to the SO splittings
quoted in the graphite literature\cite{review}.

The Rashba interaction due to a perpendicular electric field $E_z$ may be
estimated as $\lambda_R = \hbar v_F
e E_z/(4mc^2)$.  For $E_z \sim 50{\rm V}/300{\rm nm}$\cite{kato} this gives $\lambda_R
\sim .5 {\rm mK}$.  This is smaller than $\Delta_{so}$ because
$E_z$ is weaker than the atomic scale field.
The Rashba term due to interaction with a substrate is
more difficult to estimate, though since it is
presumably a weak Van der Waals interaction, this too can be expected
to be smaller than $\Delta_{so}$.

\begin{figure}
 \centerline{ \epsfig{figure=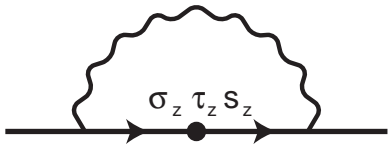,width=2in} }
 \caption{Feynman diagram describing the renormalization of the SO potential by the
 Coulomb interaction.  The solid line represents the electron propagator and the wavy line
 is the Coulomb interaction.}
 \end{figure}

This estimate of $\Delta_{so}$ ignores the effect of electron electron
interactions. The long range Coulomb interaction may
substantially increase the energy gap.  To leading order the SO
potential is renormalized by the diagram shown in Fig. 3, which
physically represents the interaction of electrons with the
exchange potential induced by $\Delta_{so}$.  This is similar in
spirit to the gap renormalizations  in 1D Luttinger liquids and
leads to a logarithmically divergent correction to
$\Delta_{so}$.  The divergence is due to the long range $1/r$
Coulomb interaction, which persists in graphene even accounting
for screening\cite{guinea}.  The divergent corrections to $\Delta_{so}$
as well as similar corrections to $\hbar v_F$ can be
summed using the renormalization group (RG)\cite{guinea}.  Introducing
the dimensionless Coulomb interaction $g = e^2/\hbar v_F$ we
integrate out the high energy degrees of freedom with energy
between $\Lambda$ and $\Lambda e^{-\ell}$.  To leading order in
$g$ the RG flow equations are
\begin{equation}
dg/d\ell = -g^2/4;\quad d\Delta_{so}/d\ell = g\Delta_{so}/2.
\end{equation}
These equations can be integrated, and at energy
scale $\varepsilon$, $\Delta_{so}(\varepsilon) = \Delta_{so}^0 [1 + (g^0/4)
\log(\Lambda^0/\varepsilon)]^2$.  Here $g^0$ and $\Delta_{so}^0$ are the
interactions at cutoff scale $\Lambda^0$.
 The renormalized  gap is
determined by $\Delta_{so}^R\sim\Delta_{so}(\Delta_{so}^R)$.   Using an effective
interaction $g^0 = .74$\cite{kanemele} and $\Lambda^0 \sim 2$ eV this
leads to $2\Delta_{so}^R \sim 15^\circ$K.

In summary, we have shown that the ground state of a single plane of
graphene exhibits a QSH effect, and has a non trivial topological order that is robust
against small perturbations.
The QSH phase should be observable by studying low temperature charge transport and
spin injection in samples of graphene with
sufficient size and purity to allow the bulk energy gap to manifest itself.
It would also be of interest to find other materials with stronger
SO coupling which exhibit this effect, as well as possible
three dimensional generalizations.

We thank J. Kikkawa and S. Murakami for helpful discussions.  This
work was supported by the NSF under MRSEC grand DMR-00-79909 and the
DOE under grant DE-FG02-ER-0145118.

\end{document}